\documentclass[lettersize,journal]{IEEEtran}
\usepackage{amsmath,amsfonts}
\usepackage{algorithmic}
\usepackage{algorithm}
\usepackage{array}
\usepackage[caption=false,font=normalsize,labelfont=sf,textfont=sf]{subfig}
\usepackage{textcomp}
\usepackage{stfloats}
\usepackage{url}
\usepackage{verbatim}
\usepackage{graphicx}
\usepackage{cite}
\def\BibTeX{{\rm B\kern-.05em{\sc i\kern-.025em b}\kern-.08em
    T\kern-.1667em\lower.7ex\hbox{E}\kern-.125emX}}
\usepackage{balance}
\begin{document}

\title{Herd Routes: A Preventative IoT-Based System for Improving Female Pedestrian Safety on City Streets}

\author{Madeleine Woodburn, Wynita M. Griggs, \IEEEmembership{Member, IEEE}, Jakub Mare\v{c}ek, \IEEEmembership{Member, IEEE} and Robert N. Shorten, \IEEEmembership{Senior Member, IEEE}
\thanks{M. Woodburn and R. N. Shorten are with the Dyson School of Design Engineering, Imperial College London, South Kingston, UK. {\tt\small madeleine.woodburn18@imperial.ac.uk}, {\tt\small r.shorten@} {\tt\small imperial.ac.uk}}
\thanks{W. M. Griggs is with the Department of Civil Engineering and the Department of Electrical and Computer Systems Engineering, Monash University, Clayton, Victoria, 3800, Australia. {\tt\small wynita.griggs@monash.edu}}
\thanks{J. Mare\v{c}ek is with the Czech Technical University in Prague, the Czech Republic. {\tt\small jakub.marecek@fel.cvut.cz}}}

\maketitle

\begin{abstract}
Over two thirds of women of all ages in the UK have experienced some form of sexual harassment in a public space. Recent tragic incidents involving female pedestrians have highlighted some of the personal safety issues that women still face in cities today. There exist many popular location-based safety applications as a result of this; however, these applications tend to take a reactive approach where action is taken only after an incident has occurred. This paper proposes a preventative approach to the problem by creating safer public environments through societal incentivisation. The proposed system, called ``\textit{Herd Routes}'', improves the safety of female pedestrians by generating busier pedestrian routes as a result of route incentivisation. A novel application of distributed ledgers is proposed to provide security and trust, a record of system users' locations and IDs, and a platform for token exchange. A proof-of-concept was developed using the simulation package SUMO (Simulation of Urban Mobility), and a smartphone app. was built in Android Studio so that pedestrian Hardware-in-the-Loop testing could be carried out to validate the technical feasibility and desirability of the system. With positive results from the initial testing of the proof-of-concept, further development could significantly contribute towards creating safer pedestrian routes through cities, and tackle the societal change that is required to improve female pedestrian safety in the long term.
\end{abstract}


\section{Introduction}
\IEEEPARstart{F}emales of all ages face gender-inequities in every day life, and the associated feelings of compromised safety and fearfulness that can arise. In cities, for instance, the day-to-day manifestation of these inequities typically centre around harassment in public spaces. Of course, in these situations, women do as much as they can to prioritise their personal safety. Notably, women approach walking through cities with extreme caution, especially at night.

In London, for example, there are ongoing initiatives such as the UN Women's Global initiative of ``Safe Cities and Safe Public Spaces for Women and Girls'', which commits to identifying gender-responsive, locally relevant and owned interventions \cite{Safe-cities-and-spaces}. Despite this, a report by the All-Party Parliamentary Group (APPG) for UN Women UK, which investigated sexual harassment in public spaces, found that 71\% of women in the UK have experienced some form of sexual harassment in a public space \cite{appg-report}. These statistics are alarming. Unfortunately, the gravity of this societal issue only reaches the public domain when there is a high profile tragedy, such as the murders of Sarah Everard, Sabina Nessa and Ashling Murphey \cite{sarah-everard,sabina-nessa,ashling-murphy} -- all happening in two years since the project's inception.

In an effort to counter the situation of potential danger walking through a city alone, especially at night, we propose an IoT-based system that incentivises people to cluster together on common walking routes. We call the system, ``\textit{Herd Routes}'', and describe its design, and demonstrate its utility. Initially, stakeholder interviews and consequence scanning workshops were employed to outline the design requirements for the system; see Section \ref{s3a}. Simulation of the system and a prototype application are then discussed in Section \ref{s3c}. The results of the prototype testing to validate the technical feasibility, and user-centred hardware-in-the-loop testing to validate the desirability, are discussed in Sections \ref{s4} and \ref{s5}. Finally, conclusions and directions for future work are presented in Section \ref{s6}.

\section{Background}\label{s2}

\subsection{Related Work}

\textbf{Personal Safety \& Internet of Things (IoT):} Within the scientific community, there are only few peer-reviewed papers specifically focused on female safety. Reference \cite{AKRAMFemaleIOT} outlines the design of a Smart Safety Device for women that relies on providing security via a fingerprint-based method of connectivity; see \cite{wifibasedconnectivity}. Although a very subtle design, it still places an expectation on the user to interact with the object, placing responsibility on them. This solution can be categorised as a reactive device, whereby action is taken only after an incident has occurred. Similarly, \cite{AKRAMFemaleIOT} cites SMARISA: A Raspberry Pi Based Smart Ring for Women Safety Using IoT \cite{SMARISA} as a reference for an IoT-based product that is embodied as a wearable and requires minimal human-interaction to instantaneously connect to the concerned authorities when a ``trigger'' movement is detected. The proposed technique uses GPS tracking of the Smart Ring to establish the user's coordinates, which can then be sent to user-specified close contacts. In summary, female pedestrian safety is an insufficiently researched area, with a wide scope existing for the contribution of a novel solution that adopts an alternative approach by using technologies at the forefront of Smart City and IoT research.

\textbf{Distributed Ledger Technologies (DLTs):} One of the key barriers in developing a safety system is establishing trust \cite{trustbasedalgo}. Recently, \cite{trustDLTframework} proposed a framework for IoT ecosystems that embeds a greater sense of trust within the architecture of the system. It also highlights some of the existing flaws of centralised systems, where trust is hard to manage, or it is not transparent where it lies, such as in the female safety applications that are currently available on the market. The paper proposes that trustworthiness be factored into all levels of system architecture and design. The Trust Reputation model suggested utilises a distributed ledger layer that manages reputation, interaction, ID management and access control. The consideration of trust is expanded to a user-centric model for access control, as proposed by \cite{hashemi2017decentralized}.

The authors of \cite{pietrosmartcities} state that current DLT applications are typically focused on payments and record keeping; however, there are plenty of unexplored applications that would provide value to our society. One of the primary benefits of implementing DLTs within a safety system is their ability to enforce compliance in sharing economy applications, which could be applied to orchestrating social behavioural change in the context of female safety within Smart Cities.

\textbf{Smart Cities, DLTs and Current Use Cases:} London is acknowledged as one the largest Smart Cities in the world, with investment specified in the Mayor of London's ``Smarter London Together'' roadmap \cite{london-smart-city}. Recently, there has been a noticeable interest in the application of DLTs within the context of Smart Cities (see Table \ref{tab:dltresearch}). A Data Verification System for CCTV Surveillance Cameras Using Blockchain Technology in Smart Cities \cite{data-verification-CCTV-system} outlines a Blockchain-based system to guarantee trustworthiness of CCTV recordings. This is relevant as it discusses privacy, immutability, and how to ensure that personal data cannot be manipulated, which will be a focus point in the proposed female safety system. Overall, the importance of authentication and verification of data for Smart Cities will be a critical consideration for this project.  Current research has proven that DLTs and IoT can work together to solve many of the existing security issues and concerns.

Vehicular networks within Smart Cities have been further explored in Block-VN: a distributed Blockchain based vehicular network architecture in Smart Cities \cite{Block-vn}, where a secure environment is proposed for both end-users and the machine side. The concept of wearables used in this network was explored by \cite{griggs-localising-missing-entities}. The paper outlines a modelling technique that uses Simulation of Urban Mobility (SUMO) software \cite{sumo-software}. SUMO has proved to be an effective, open source technique for modelling wide-scale urban systems and making iterative changes, and therefore will be similarly used to model the proposed outcome for this project.

\subsection{Current Market Solutions}

There is market saturation in applications that adopt a reactive approach to the problem under consideration. Applications such as Safe \& The City \cite{safe-and-the-city}, Path Community \cite{path-community} and Safetipin \cite{safetipin} set themselves apart from the market by how they share data about unsafe experiences. In creating a database of crowd-sourced safety reports and alerting authorities such as police and local councils if any incident occurs, they aim to reduce incidents and user fear.

CityMapper\footnote{\url{https://citymapper.com}}, a popular route navigation application specifically designed for use in cities, recently released a ``main road'' safety feature which acknowledges that ``Sometimes, the fastest way to get there isn't the best way to get there'' \cite{citymapper-main-roads}. This was implemented after female users had complained that the routes suggested by the application's algorithm had led them through badly lit, quiet streets with multiple turns -- and even through parks late at night. This feature has been successful as females seeking a greater feeling of safety prefer to walk down busier, well-lit main roads than the alternative fastest route.

\subsection{Opportunity Area, and Novel Contribution of the Paper}

It is clear that there is a great opportunity to apply the learnings from current DLT use cases within Smart Cities to the issue of female pedestrian safety. There is scope to develop a system that is truly preventative, where action is taken before any incident occurs and actively seeks to create safer environments rather than just reporting data post-incident.

The issue of female pedestrian safety affects, directly or indirectly, all citizens. Therefore, all citizens should be responsible for solving it. By designing a solution that does not solely rely on the altruism of others, but provides individual financial incentives, a more realistic system is likely to become an adopted, effective solution. Of course, a system design is a short term contribution to an issue that needs to be addressed over a longer period. This will require education around the topic area, ensuring that it is a problem that is solved together, not one where the onus of solving it falls on the victim.

The success of the solution presented in this paper has been quantified using a combination of modelling results and user-feedback. Three key considerations have been used to determine the metrics of success while validating the feasibility of the project:
\begin{enumerate}
  \item Security -- ensuring that the safety of the system users is prioritised;
  \item Effectiveness -- actively improving the safety of vulnerable pedestrians; and
  \item Inclusivity -- inclusive of all minority groups of vulnerable pedestrians. The system is scalable to various contexts within different cities.
\end{enumerate}

\begin{table*}[!h]
\caption{Existing research on DLT applications in Smart Cities, applicable to user data and anonymity. \label{tab:dltresearch}}
\centering
\begin{tabular}{|c|c|c|c|c|}
\hline
\textbf{Reference} & \textbf{Year} & \textbf{Ledger platform} & \textbf{Data} & \textbf{Research overview}\\ [0.5ex]
\hline
\cite{ethereum-medical-data} & 2016 & Ethereum & Medical data & Immutable patient medical data\\
\hline
\cite{hashemi2017decentralized} & 2017 & Bitcoin & User data & User-centric control of personal data\\
\hline
\cite{hyperledger-fabric-anonymised} & 2017 & Hyperledger Fabric & Anonymised personal data  & Data transfer between broker and receiver\\
\hline
\cite{Block-vn} & 2017 & Block-VN & Vehicle information & Distributed network of vehicles in a Smart City\\
\hline
\cite{hyperledger-prescription-drug} & 2019 & Hyperledger Fabric & Prescription drugs record & Integrity management for hospitals\\
\hline
\end{tabular}
\end{table*}

\section{Methodology}

\subsection{Problem Area Exploration}\label{s3a}

\textbf{Stakeholder Interviews:}
To design and  implement a solution that will be successfully adopted by its various stakeholders, interviews were conducted to gain deeper insight into the requirements of the system. Four value-oriented, semi-structured interviews were held. Due to their semi-structured nature, the questions posed acted more like prompts so that the conversation could follow a natural course. This method sought to gain the most insight from the interaction, and remove the opportunity for bias in the questioning.

\textbf{Consequence Scanning Workshop:}
The problem area is multi-faceted and has many ethical concerns that needed to be explored further by utilising an investigative method such as Consequence Scanning \cite{consequence-scanning-toolkit}. The Consequence Scanning workshop was conducted on Figma \cite{figma}, a collaborative online tool. The participants were not familiar with this type of workshop; and therefore, it was adapted to be more accessible and easily understood. The workshop was split into three sections: identifying intended and unintended consequences of the suggested solution; prioritisation of these consequences with the appropriate measure taken; and finally, how to address the consequences with design requirements.

\subsection{Development \& Iteration of Concepts
}\label{s3b}

\textbf{Divergent Concepts:}
The first concept, ``Paired Routes'', allowed two system users' routes to be matched together and then superimposed to maximise the amount of time that they were walking together. Each interaction between system users would be stored on a distributed ledger. The second concept was a token exchange concept which allowed a user to request a `buddy' that would escort them along their route. The token exchange would incentivise the `buddies' to escort users to their destination, similar to an Uber ride-hailing model. After discussions with stakeholders, these concepts were discarded as the risk of system misuse would be too high, and would need to rely on thorough and heavily secured verification of new users.

\textbf{Selected Concept -- \textit{``Herd Routes''}:}
The third concept, entitled ``Herd Routes'', was chosen as the selected concept. As opposed to the discarded ones, this system proposal has no need for interaction with another system user. It also utilises distributed ledgers as a way of incentivising societal behaviour change \cite{pietrosmartcities}, which is an integral part of the long-term vision for creating an environment where women are not subject to sexual harassment in public spaces.

The concept relies on the insight that busier streets have a greater perceived safety for women. This logic was similarly applied in CityMapper's ``Main Road'' feature: when deciding between an empty street and a busier one, women are more likely to choose the latter to walk along. Herd Routes generates pedestrian flow along specified routes by incentivising system users with payment in the form of tokens. As well as increasing safety by generating busier routes, the system should also keep a record of which system users were on the routes over time, similar to other navigation-based safety applications. This would not necessarily increase perceived safety in the instant, but would provide evidence if an incident were to occur.

\begin{figure*}[!t]
\centering\includegraphics[width=2.00\columnwidth]{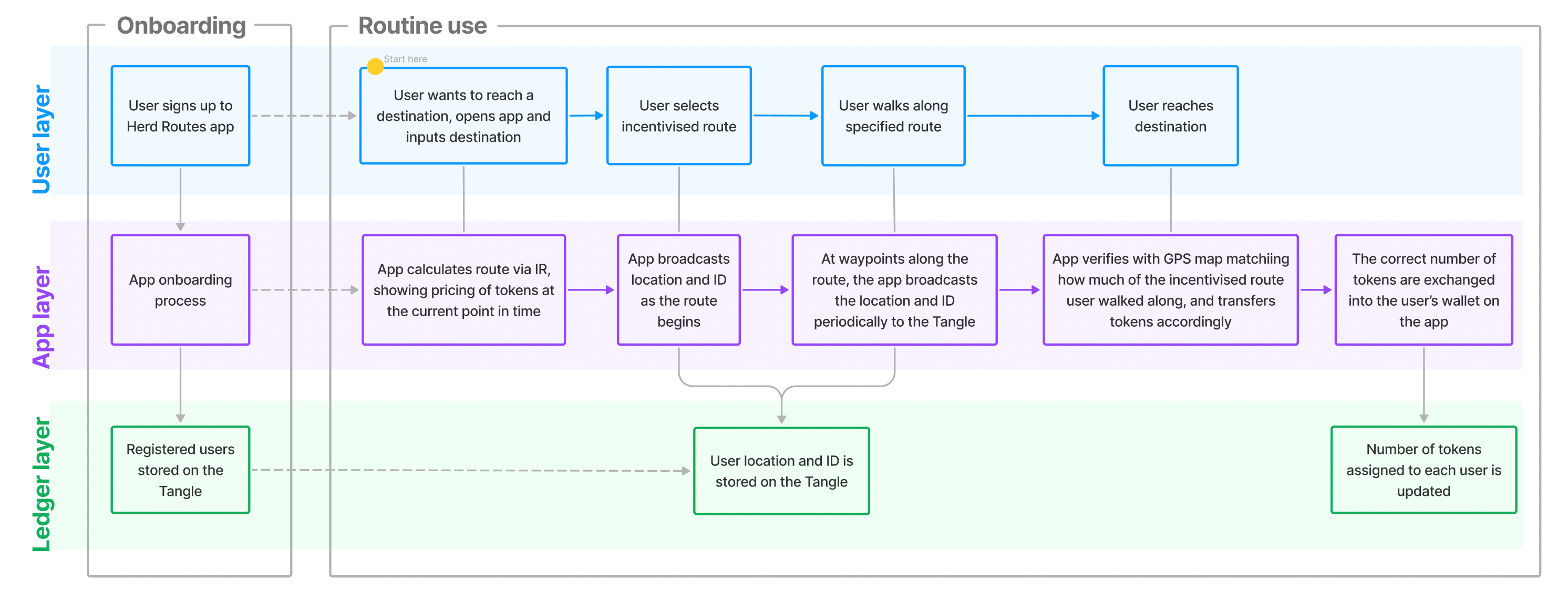}
\caption{Herd Routes system overview.
}
\label{System-overview}
\end{figure*}

The user inputs their final destination into the application, and the algorithm routes them along the incentivised route, which would be busier due to the generated pedestrian flow. This is a system that all citizens (even those who do not feel unsafe) can use, and therefore contribute towards creating safer public spaces for female pedestrians. The user-facing embodiment of this system will be a route navigation application. As seen in Fig. \ref{System-overview}, the system can be thought of in three distinct layers: the user; the application; and the ledger. All three layers interact with each other to create a holistic view of the system.

The security, trust and core functionality of the Herd Routes system is facilitated by DLTs. As shown in Section \ref{s2}, the various use cases and functionalities of DLTs in Smart Cities have proven to be very successful. There are, however, many different types of platforms, the most notable of which being Blockchain\cite{blockchain-website}. Although widely used, this platform has drawbacks that mean it is not viable for use in the IoT industry. Blockchain has an associated transaction fee for any value of transaction, meaning micropayments become illogical if the transaction fee is greater than the value of payment. Due to the fact that Blockchain participants can be categorised into those who transact and those who approve transactions, an opportunity presents for inadvertent discrimination or misuse of the system. For the Herd Routes system, using a ledger platform that is inherently discriminatory works against many ethical design considerations \cite{Tangle-popov}. The IOTA Tangle\footnote{\url{https://www.iota.org/get-started/what-is-iota}} was developed specifically for use within the IoT industry, due to its high scalability, zero fees, and near-instant transfers, and therefore was selected as the ledger that best supported this system \cite{iota-tangle-dag}.

The Herd Routes system employs a dynamic pricing controller to price the tokens according to how many system users there currently are and how many there needs to be for the street to be perceived as busy. As a result of this, the system forms a positive feedback loop, where the number of system users, the token price, and the number of non-system users all influence one another.

\subsection{Build \& Test}\label{s3c}

\begin{figure*}[b]
\centering\includegraphics[width=2.00\columnwidth]{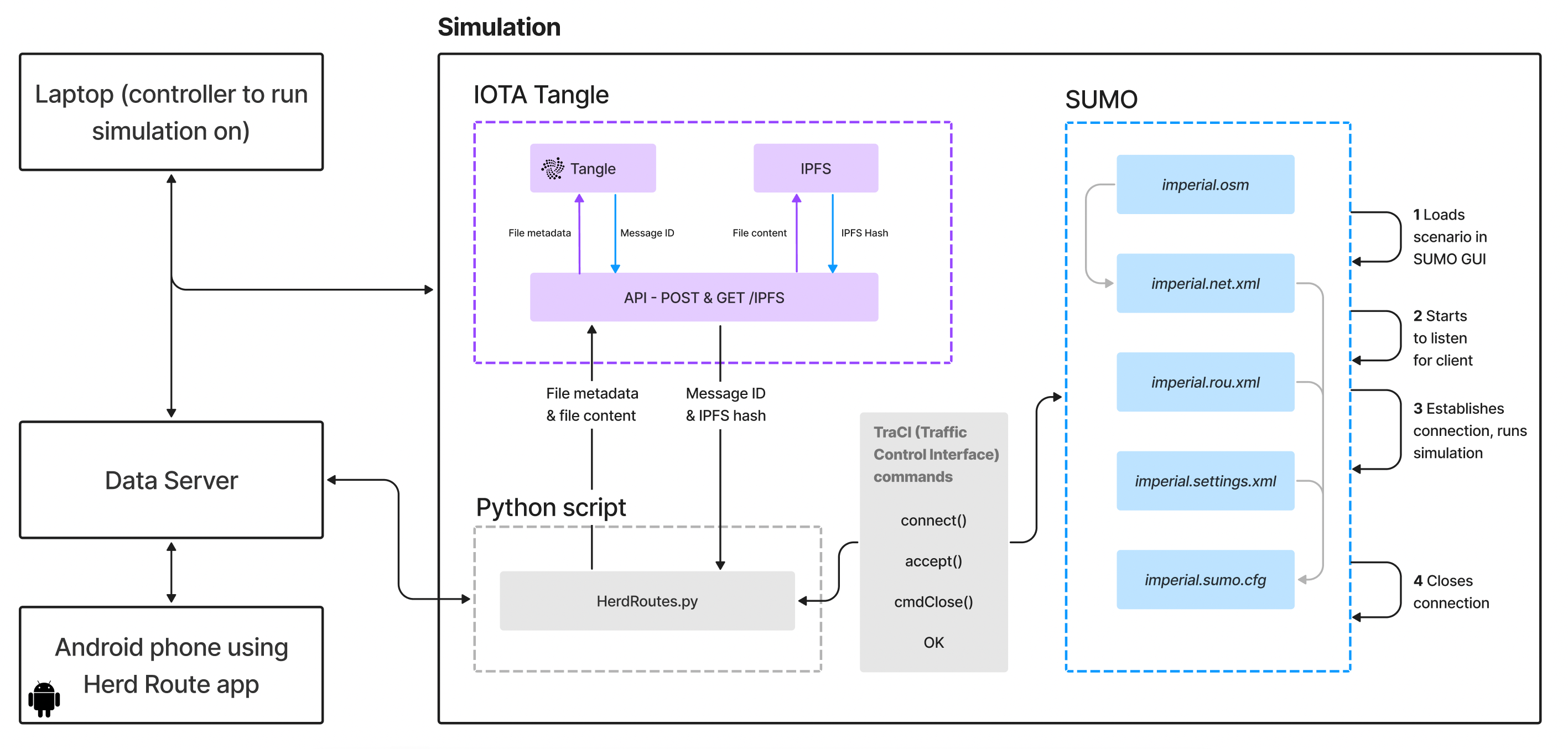}
\caption{System architecture for the prototype, used in the Hardware-in-the-Loop testing.}
\label{tech-system-architecture}
\end{figure*}

\textbf{Prototype Overview:}
Herd Routes is a large-scale urban system that could not be fully prototyped at this early stage of its conception due to its complexity. Therefore, to validate the concept and explore the feasibility of the system, a model was built to replicate its core functionalities. The build of the prototype was broken down into two phases: Phase (1) the development of a simulation of the system using SUMO software and Python; and Phase (2) a Hardware-(or Person-)in-the-Loop test that utilised this simulation. In particular, once the simulation of the system was constructed, an Android smartphone application, for a real person to use, was designed and built for integration with the simulation. The smartphone application communicated with the simulation via an external server. The system architecture for the prototype (see Fig. \ref{tech-system-architecture}) shows how the Python script interacted with the SUMO software, posted message data to the Tangle, and received incoming data from the Android device.

\textbf{Phase (1) -- SUMO Simulation:} SUMO is an open-source, multi-mode simulation package that has been designed to model large-scale urban networks. As mentioned in existing literature, it is widely used in industry and research due to its flexibility and wealth of resources and documentation \cite{sumo-software}. The SUMO simulation was developed in the Visual Studio Code IDE using Python 3.9.7. The Python script communicated with the SUMO software using Traffic Control Interface (TraCI) APIs. TraCI uses a TCP-based client/server architecture to provide access to SUMO. Once a connection had been established, the client could edit and manipulate the simulation, request information about any agent or environment, and ultimately close the connection down.

\begin{figure}[!b]
\centering\includegraphics[width=0.95\columnwidth]{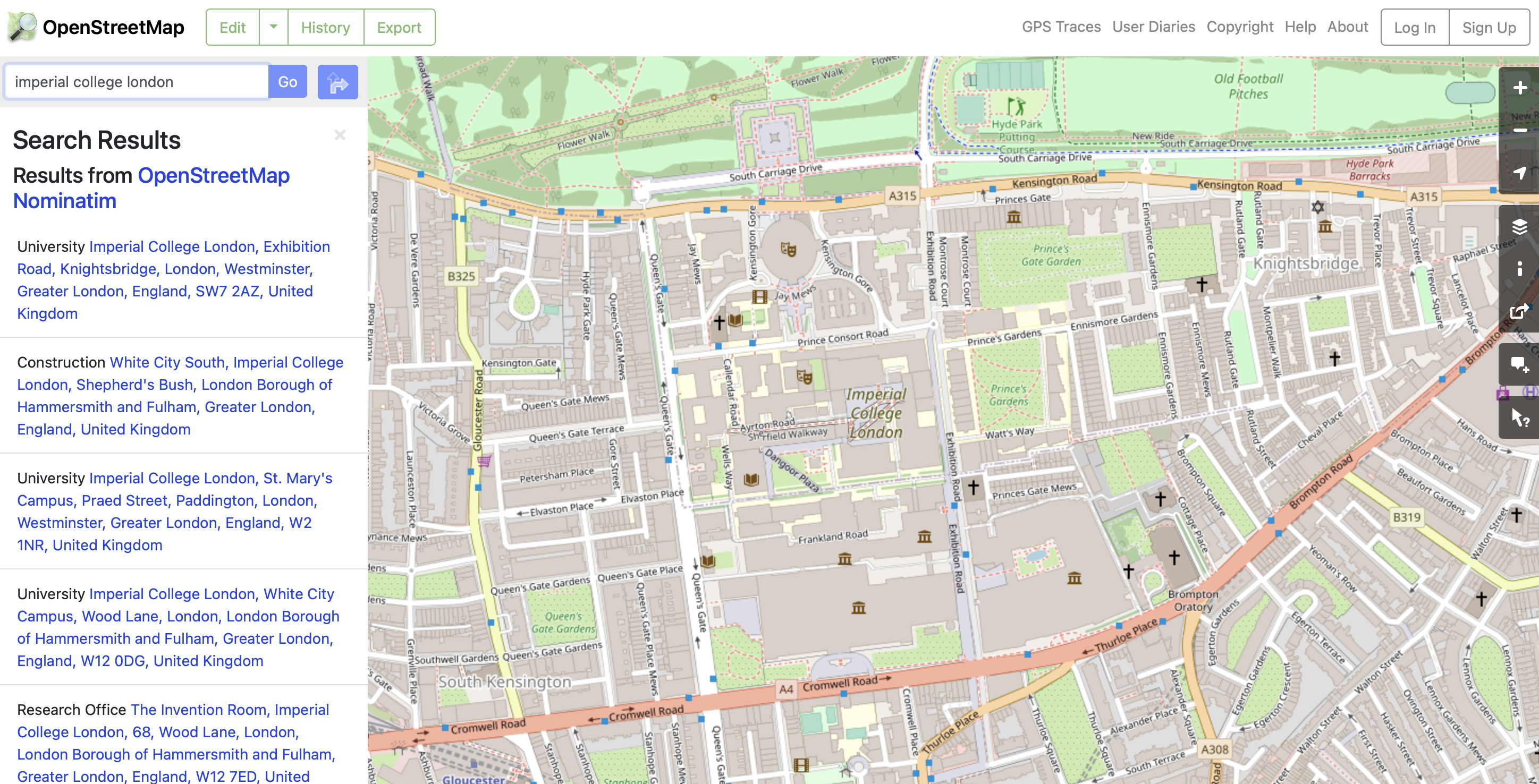}
\caption{Downloaded area from OpenStreetMap.}
\label{osm-download}
\end{figure}

\begin{figure}[!b]
\centering\includegraphics[width=0.95\columnwidth]{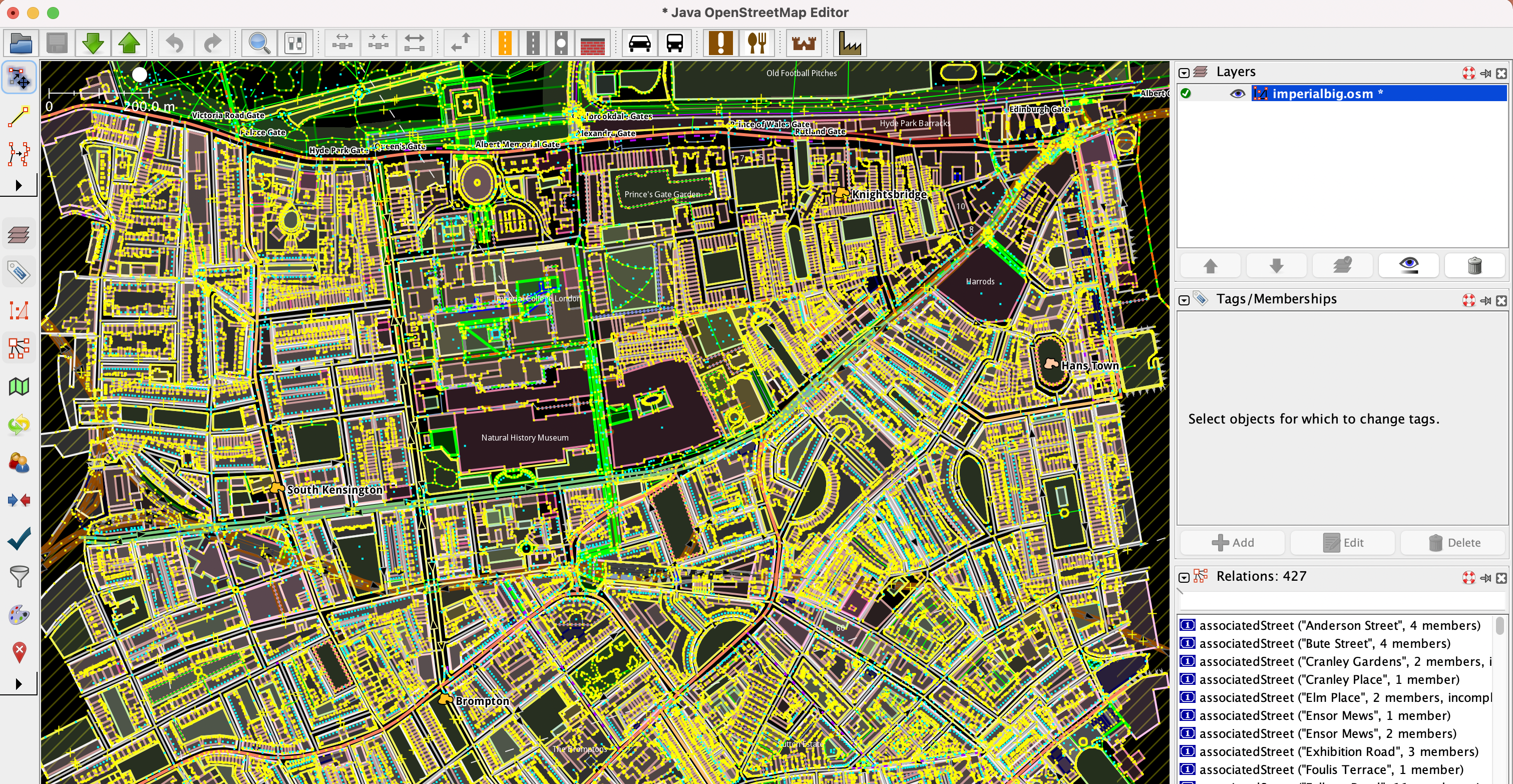}
\caption{The map was edited, removing unnecessary features and making sure all roads had the correct pedestrian properties.}
\label{josm-edit}
\end{figure}

The SUMO software also extends to a graphical user interface, called SUMO-GUI. It provided useful visualisation of the simulated environment, displaying the movement of agents through the map. For the purpose of the Hardware-in-the-Loop testing, a map of the Imperial College South Kensington campus was downloaded from OpenStreetMap\footnote{\url{https://www.openstreetmap.org/#map=15/51.4974/-0.1776}} (Fig. \ref{osm-download}), edited (Fig. \ref{josm-edit}) and then inputted into the SUMO network generation command. The agent visualisation was useful while the simulation was being developed and iterated, as it provided a sanity-check for modelling the rationale of the agents, ensuring that the functionality of the simulation was as intended.

The main algorithm (Algorithm \ref{alg1}) used for the simulation was comprised of three core functionalities, which were implemented as a further three discrete algorithms:
\begin{enumerate}
  \item incentivisation of certain routes (Algorithm \ref{alg2});
  \item dynamic pricing control (modelling rational human behaviour) (Algorithm \ref{alg3}); and
  \item IOTA communications and protocols (Algorithm \ref{alg4}).
\end{enumerate}
Rational pedestrian agents were modelled, added to the system, and assigned a random route. The walking stages of the route were then appended to their path within the simulation. Every 10 simulation time steps, the token price was updated. If the token price had changed compared to the previous price that was set 10 time steps ago, then the agent would re-decide whether to accept the new token price and thus utilise an incentivised route, or not. Once on an incentivised route, and at each waypoint along it, the agent would post their location and system user ID to the IOTA Tangle, thus paving the way for the pedestrian to be paid for walking along an incentivised route. Further details can be found in the pseudocode for Algorithms \ref{alg1} to \ref{alg4}.

\begin{algorithm}[H]
\caption{SUMO Simulation Main Algorithm}
\begin{algorithmic}
\STATE
\STATE {\textbf{Data:}} Number $N$ of agents; \text{number $T$ of simulation steps}; fixed interval $I$ for token-pricing updates.
\STATE {\textbf{Result:}} Specified number $N$ of pedestrians are inserted to the simulation to validate the dynamic pricing, incentivisation of routes, and IOTA checkpoint elements of the Herd Routes system.
\STATE
\vspace{0.2cm}
// \textit{Add agents into simulation with routes on permitted lanes.}
\STATE \textbf{Generate} $N$ simulated agents
\STATE \textbf{Create} connection to TraCI
\vspace{0.2cm}
\STATE \textbf{for} each lane in the road network that permits pedestrians
\STATE \hspace{0.4cm} append lane to {\tt pedestrianAllowedLanes} list
\STATE \textbf{end}
\vspace{0.2cm}
\STATE \textbf{for} each simulated agent
\STATE \hspace{0.4cm} add pedestrians
\STATE \hspace{0.4cm} add route utilizing  {\tt pedestrianAllowedLanes}
\STATE \textbf{end}
\vspace{6pt}
\STATE // \textit{Determine which agents initially decide to take}
\STATE // \textit{an incentivised route based on initial token pricing.}
\STATE Set initial token price
\STATE \textbf{for} each simulated agent
\STATE \hspace{0.4cm} Let the agent decide
\STATE \hspace{0.4cm} \textbf{if} agents decides `yes'
\STATE \hspace{0.8cm} append agent ID to list of agents initially taking an
\STATE \hspace{1.2cm} incentivised route
\STATE \hspace{0.8cm} select incentivised route
\STATE \hspace{0.8cm} navigate agent onto selected
\STATE \hspace{1.2cm} incentivised route
\STATE \hspace{0.4cm} \textbf{end}
\STATE \textbf{end}
\STATE
\vspace{6pt}
// \textit{Run the main part of the simulation.}
\STATE \textbf{while} $i < T$ :
\STATE \hspace{0.4cm} \textbf{if} $i$ is divisible by $I$:
\STATE \hspace{0.8cm} update the token price using \textbf{Algorithm \ref{alg3}}
\STATE \hspace{0.8cm} \textbf{for} agents who decided to use an incentivised route:
\STATE \hspace{1.2cm} set the price using \textbf{Algorithm \ref{alg2}}
\STATE \hspace{0.8cm} \textbf{end}
\STATE \hspace{0.4cm} \textbf{end}
\STATE \hspace{0.4cm} $i$ +=1
\vspace{0.2cm}
\STATE \textbf{Close} connection to SUMO
\end{algorithmic}
\label{alg1}
\end{algorithm}

\begin{figure}[!b]
\centering\includegraphics[width=0.8\columnwidth]{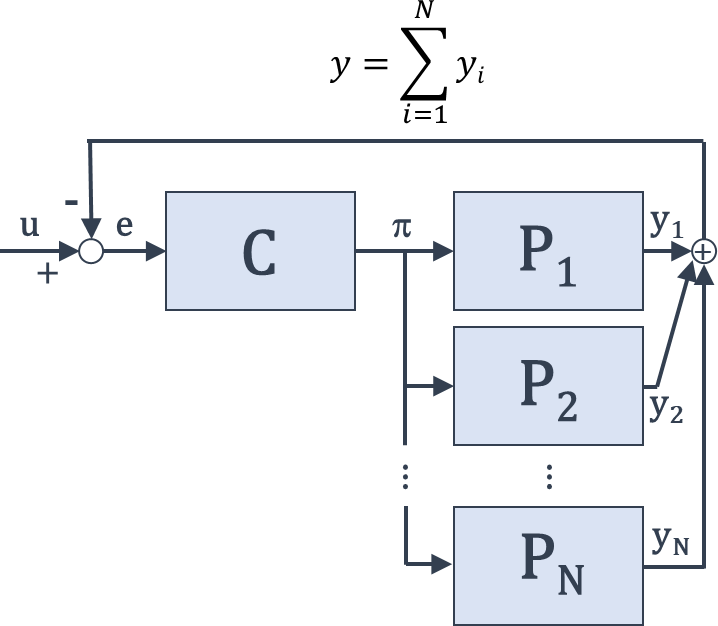}
\caption{Dynamic pricing feedback loop to set the value of tokens.}
\label{pid-controller}
\end{figure}

\textit{Dynamic Price Control:} Dynamic pricing control was applied to regulate the density of people along incentivised routes. The dynamic price control was implemented using a proportional integral derivative (PID) controller such that the updated price of the tokens was set according to the values of the prior and present error signals (where an error signal, $e$, is the difference between the desired number of system users on the incentivised routes and the actual number of system users on the incentivised routes); and the prior control output, $\pi$. Mathematically, the PID controller was specified by \begin{equation}\label{e1}
    \pi(k) = \beta\pi(k-1)+\kappa[e(k)-\alpha e(k-1)]
\end{equation}
where $\alpha = -4.01$, $\beta=0.99$ and $\kappa=0.1$; and is also indicated in Fig. \ref{pid-controller} by $C$. Moreover, in Fig. \ref{pid-controller}:
\begin{itemize}
  \item \textbf{$C$} denotes the controller which takes the error, $e$, as input and produces a price, $\pi$, which is the payment to be offered to each person to entice them to use an incentivised route;
  \item \textbf{$P_i$} denotes a person who is either using an incentivised route\footnote{For simplicity, in this paper, a set of incentivised routes was created and, once deciding to use an incentivised route, each agent then chose a random incentivised route from the set. That is, the distance to each incentivised route was not considered. This is an additional complexity that could be factored into future work.} or not;
  \item \textbf{$y$} is the total number of people currently using an incentivised route, and is a random number;
  \item \textbf{$u$} is the desired number of people using an incentivised route, as stipulated by the city based on some safety criteria, and is a constant number;
  \item \textbf{$e$} denotes the error signal and is equal to $u–y$;
  \item and the parameters $\alpha$, $\beta$ and $\kappa$ were based on the values used in \cite{controller-default-values}.
\end{itemize}

\begin{figure}[!t]
\centering\includegraphics[width=0.9\columnwidth]{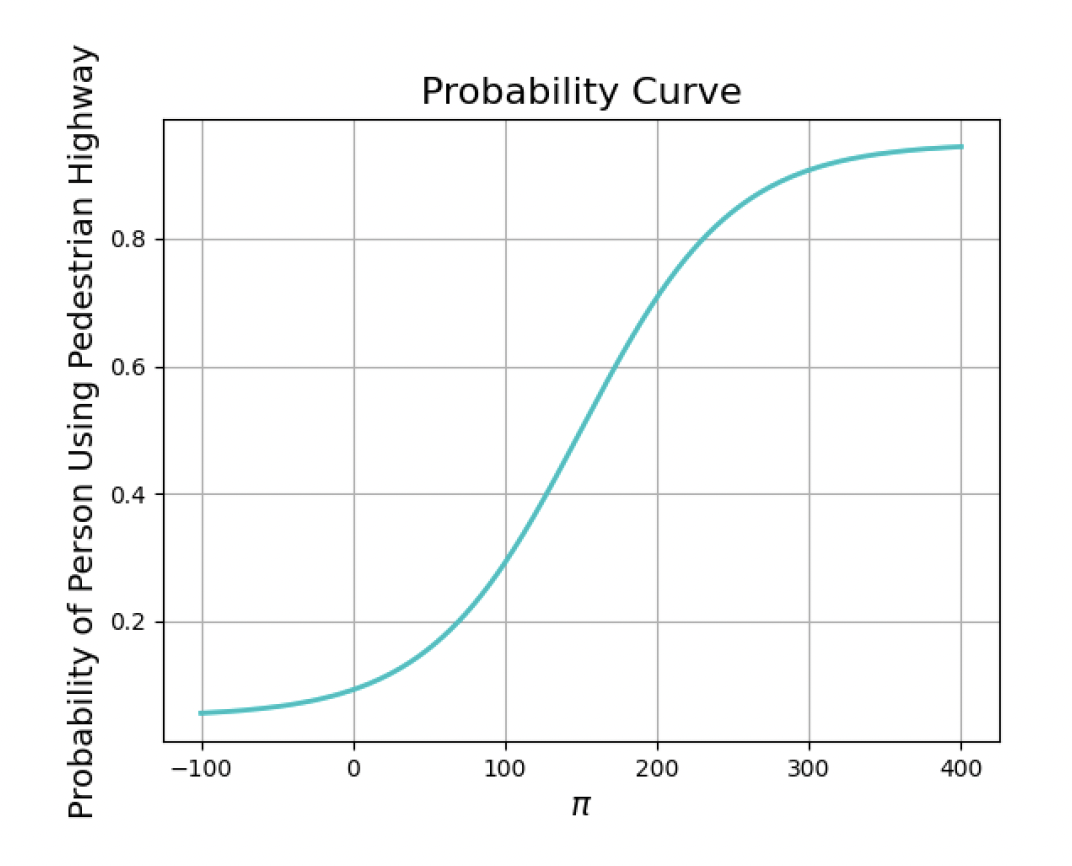}
\caption{The probability of a pedestrian using an incentivised route, versus the token price. This simple model reflects the typical price acceptance for a rational agent.}
\label{controller-probability}
\end{figure}

Each decision of whether a person rerouted to an incentivised route or not was determined as follows, where the decisions were made at the beginning of the simulation, and then every time the token price was updated. First, the probability that the person would accept the initial or updated token price, and thus re-route to an incentivised route, was found by locating the initial or updated token price $\pi$ along the x-axis of Fig. \ref{controller-probability}, and observing the probability associated with that price on the y-axis. Note that the function in Fig. \ref{controller-probability} is demonstrative of a very basic model of a rational agent: the probability that an agent will accept the token price increases as the price itself (i.e., the amount that the agent will be paid) increases. In this paper, for simplicity, all agents' decision-making was described by the same basic model. Next, for each agent, the Python function {\tt random.uniform(0,0.25)} was invoked.\footnote{Note that the upper limit value of 0.25 was chosen for illustrative purposes, and could be replaced with a value to better reflect real human behaviour if required.} Finally, the output of the uniform random number generator (which was unique for each agent) was compared with the probability obtained from the y-axis of Fig. \ref{controller-probability} (which was the same value for each agent, for simplicity, in this paper). If the output of the random number generator was less than the probability determined from the y-axis of Fig. \ref{controller-probability}, then the agent decided to use an incentivised route; and vice versa.

The dynamic pricing control algorithm (Algorithm \ref{alg3}) was tested to evaluate the suitability of the model over time. The model parameters were set such that the total population size was 750; the controller was updated every 10 time steps; and 10 simulations were run in total (with each simulation consisting of 2000 time steps), so that the convergence of the control output, and the success in regulating the feedback control loop, could be observed in terms of mean values and standard deviations. The target density of pedestrians on the incentivised routes (i.e., {\tt fixedDemand}, see Algorithm \ref{alg3}) was tuned to: (a) find an acceptable value of {\tt fixedDemand} such that some criteria would be met (e.g., people would feel safe); and (b) check that the controller worked in the extremities of possible cases.

\begin{algorithm}[H]
\caption{Dynamic Pricing Control Algorithm}
\begin{algorithmic}
\STATE
\STATE {\textbf{Data:}} {\tt fixedDemand}; Agent Population Information: agents {\tt agentPop} in the population; agents {\tt agentsOn} utilising an incentivised route; Controller Parameters: $e$, $e_{\textrm History}$, $\pi$, $\pi_{\textrm History}$, $\alpha$, $\beta$, $\kappa$.
\STATE {\textbf{Result:}} Updated token price.
\vspace{0.2cm}

\STATE $e$ = {\tt fixedDemand} -  {\tt agentsOn}
\STATE $\pi$ = $\beta$ $\times$ $\pi_{\textrm History}$ + $\kappa$[$e$ - $\alpha$ $\times$ $e_{\textrm History}$], cf. Eq. \ref{e1}
\STATE $e_{\textrm History}$ = $e$
\STATE $\pi_{\textrm History}$ = $\pi$
\vspace{0.2cm}
\STATE Reset {\tt agentsOn}
\STATE \textbf{for} agent in {\tt agentPop}:
\STATE \hspace{0.4cm} call {\tt ProbabilityFunction}($\pi$), cf. Fig. \ref{controller-probability}
\STATE \hspace{0.4cm} Let the agent decide
\STATE \hspace{0.4cm} \textbf{if} agent decides `yes':
\STATE \hspace{.8cm} append agent to {\tt agentsOn}
\STATE \hspace{0.4cm} \textbf{end}
\STATE \textbf{end}

\end{algorithmic}
\label{alg3}
\end{algorithm}

\textit{IOTA -- Location and ID Posting, and Token Exchange:} Within the Herd Routes system, the inclusion of a distributed ledger enabled three primary functions:
\begin{enumerate}
  \item a means of trust and security to the IoT system;
  \item a ledger for system user IDs and locations to be stored;
  \item a platform for token exchange after completion of an incentivised journey.
\end{enumerate}

The second functionality was demonstrated in the Python simulation, as this was the minimum functionality required for the proof-of-concept. Due to changes in IOTA protocols that were deployed in April 2021, \cite{iota-update}, Chrysalis (known as IOTA 1.5) with Hornet node software was deployed within the simulation. The Client class was used to provide high-level abstraction of the IOTA.rs library. This class was instantiated before starting any interactions with IOTA nodes. The library automatically chose a starting node within the Devnet\footnote{\url{https://legacy.docs.iota.works/docs/getting-started/1.2/networks/devnet}} (similar to the Mainnet, except tokens are free and it incurs less computational power to interface) for the simulation to connect. After the health of the network was checked, the IOTA update (see Algorithm \ref{alg4}) was called.

\begin{algorithm}[H]
\caption{Incentivised Route Algorithm}
\begin{algorithmic}
\STATE
\STATE {\textbf{Data: }}
Agent Population Information: agents {\tt agentsOn} utilising an incentivised route;
agent status (i.e., agents who have newly decided to utilise {\tt new}, as opposed to who are already utilising, an incentivised route).
\STATE {\textbf{Result: }} Reroute {\tt new} agents onto the selected incentivised route. Do IOTA Location and ID Posting for all agents utilising an incentivised route.
\STATE
\vspace{0.2cm} \textbf{if} agent in {\tt new}:
\STATE \hspace{0.4cm} select incentivised route;
\STATE \hspace{0.4cm} navigate agent onto incentivised route.
\STATE \textbf{end}
\STATE
\vspace{0.2cm}
\textbf{while} agent in {\tt agentsOn}:
\STATE \hspace{0.4cm} check location
\STATE \hspace{0.4cm} \textbf{if} agent is at a waypoint:
\STATE \hspace{0.8cm} \textbf{do} IOTA Location \& ID Posting (\textbf{Algorithm \ref{alg4}})
\STATE \hspace{0.4cm} \textbf{end}
\STATE \hspace{0.4cm} \textbf{if} agent completes incentivised route:
\STATE \hspace{0.8cm} navigate back to the original destination
\STATE \hspace{0.4cm} \textbf{end}
\STATE \textbf{end}
\end{algorithmic}
\label{alg2}
\end{algorithm}

If an agent within the simulation had navigated onto an incentivised route, their location and ID (retrieved via the TraCI commands {\tt TraCi.person.getlocation()} and {\tt Traci.person.getID()}, respectively) were sent to Devnet at each waypoint that they passed along the route. In doing so, a record of the location of each system user was stored in case of an incident (e.g., harassment or an incursion on the pedestrian's personal safety while walking).

\begin{algorithm}[H]
\caption{IOTA Location and ID Posting Algorithm}
\begin{algorithmic}
\STATE
\STATE {\textbf{Data:}} Pedestrian data (location and ID)
\STATE {\textbf{Result: }} Connect to Tangle; check the health of the Devnet; post agent's location and ID to Tangle.
\STATE
\vspace{0.2cm} Generate random seed and save it
\STATE Generate an empty address using the random seed
\STATE
\vspace{0.2cm} Format SUMO pedestrian data into a  Tangle message
\STATE Check health of Devnet and generate address.
\STATE Post formatted pedestrian data to Tangle.
\end{algorithmic}
\label{alg4}
\end{algorithm}

\begin{figure}[!b]
\centering\includegraphics[width=0.9\columnwidth]{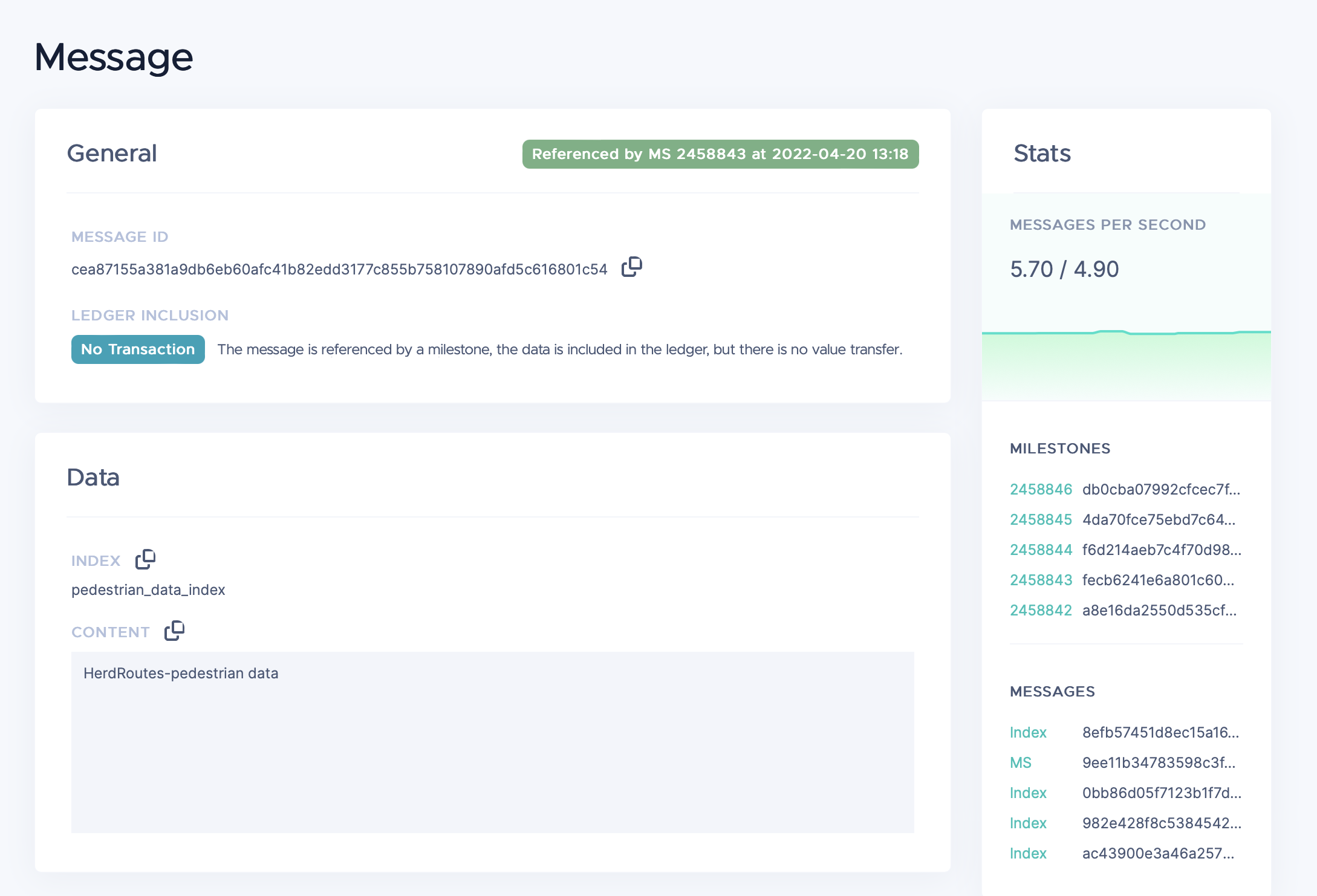}
\caption{The Tangle explorer, with an example message that the simulation sent to the Devnet.}
\label{tangle-explorer}
\end{figure}

This information was sent as a message (that is, a type of data structure that is broadcast to the IOTA network and represents a node in the Tangle graph). Every message is referenced by a {\tt message\_id} which is based on a hash algorithm of binary content. For the proof of concept, {\tt IndexationPayload} was used. This type of payload enabled the addition of an index to the encapsulating message, as well as the raw message data. For prototyping purposes, this was the most useful as it allowed for an easier search within the Tangle explorer (see Fig. \ref{tangle-explorer}), which was used to validate the completion of the data post.

\begin{figure}[!b]
\centering\includegraphics[width=0.6\columnwidth]{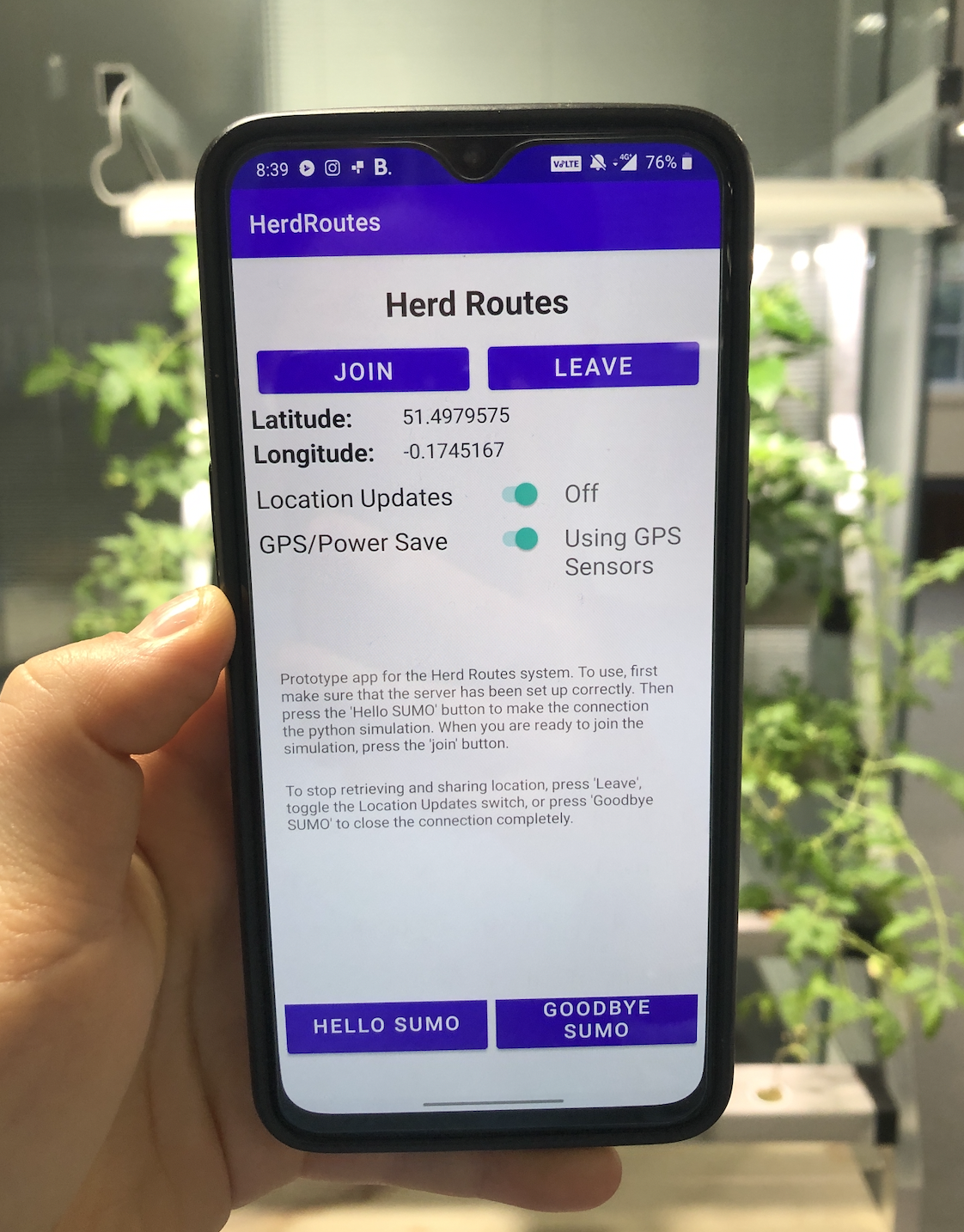}
\caption{Herd Routes prototype smartphone application downloaded on an Android device to be used in Hardware-in-the-Loop simulation.}
\label{HerdRoutes-app}
\end{figure}

\textbf{Phase (2) -- Hardware-in-the-Loop (HIL) Testing (Application Build):} Simulating the system with Python and SUMO proved useful in evaluating the functioning of the implemented pricing control and IOTA integration; however, simulation alone was not enough to gauge the perspective of a real user on the design.
Therefore, a smartphone application (see Fig. \ref{HerdRoutes-app}) was built in the native Android IDE \cite{android-studio-ide} and coded in Java and XML languages.
Using HIL testing allowed for a human-centred approach to be taken, with behaviour and responses directly inputted into the simulation. It enabled user feedback (not provided by the simulation alone), which was utilised to make iterations to the system design. We note that the HIL method has been successfully employed in other wide-scale urban system projects, such as \cite{griggs-hil-platform} and \cite{shaun-sweeney-bike}.

Google Play Services was used to retrieve the GPS data from the phone. Version 19.0.1 of {\tt play-services-location} was included in the gradle dependency list in order for the Location Provider library to be functional. This was utilised in the {\tt SendGPS} and {\tt updateGPS} methods to retrieve the user's GPS location. The application connected to the simulation by using a client socket on an external host IP address and designated port number. This allowed the transfer of the GPS coordinates in the worker thread to be processed by the Python simulation. Once processed, the GPS coordinates were converted to Cartesian X and Y coordinates by the TraCI {\tt ConvertGeo} method, which were then used to find the nearest {\tt edgeID} in the map for the real-life agent to be inserted.

\section{Evaluation \& Results}\label{s4}

\subsection{Dynamic Pricing Control}

\begin{figure}[!b]
\centering\includegraphics[width=1\columnwidth]{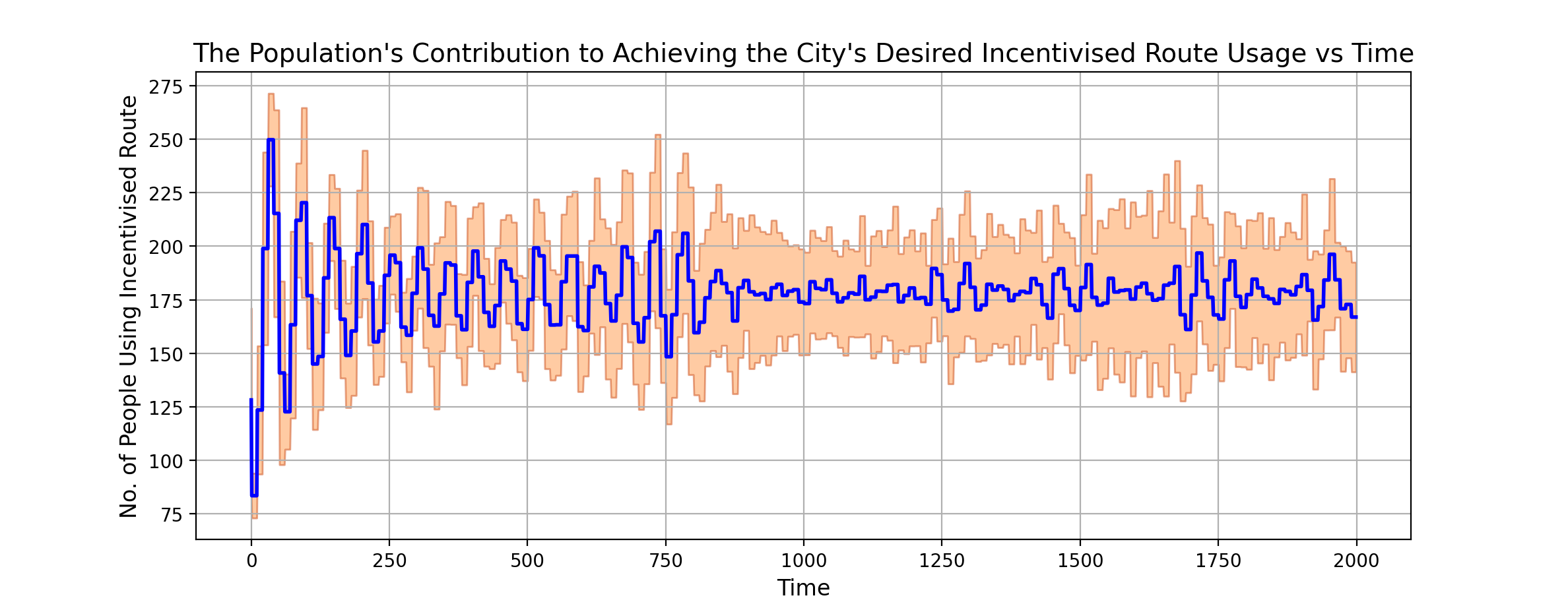}
\caption{The population's contribution to meeting the target for pedestrian highway usage. The blue line indicates the mean number of people using the route, while the red shaded area indicates one standard deviation from the mean.
}
\label{contribution-streets}
\end{figure}

\begin{figure}[!t]
\centering\includegraphics[width=1\columnwidth]{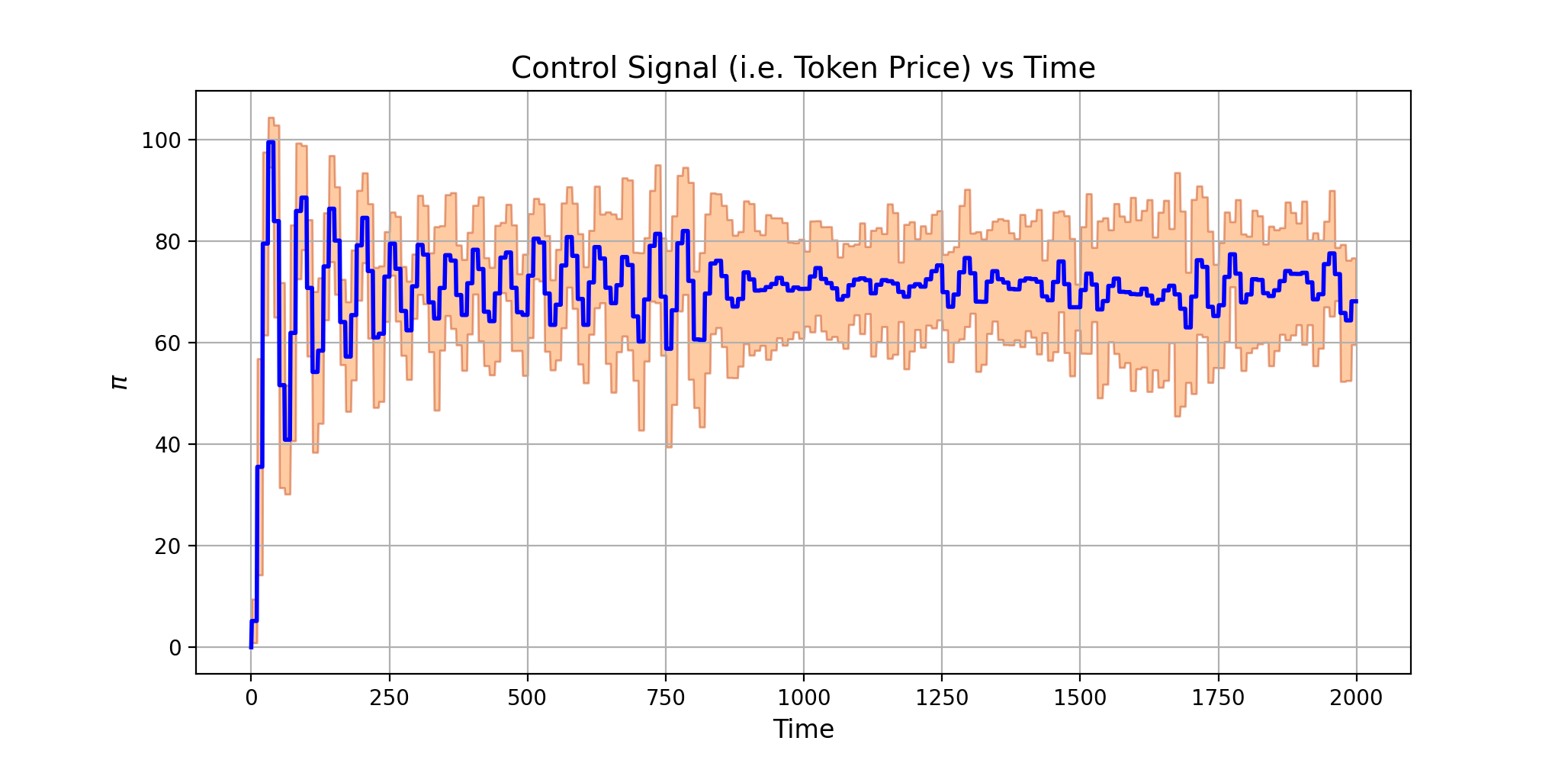}
\caption{The evolution over time of the output from the controller, $\pi$. The blue line indicates the mean of $\pi$, while the red shaded area indicates one standard deviation from the mean.}
\label{price-vs-time}
\end{figure}

\begin{figure}[!t]
\centering\includegraphics[width=1\columnwidth]{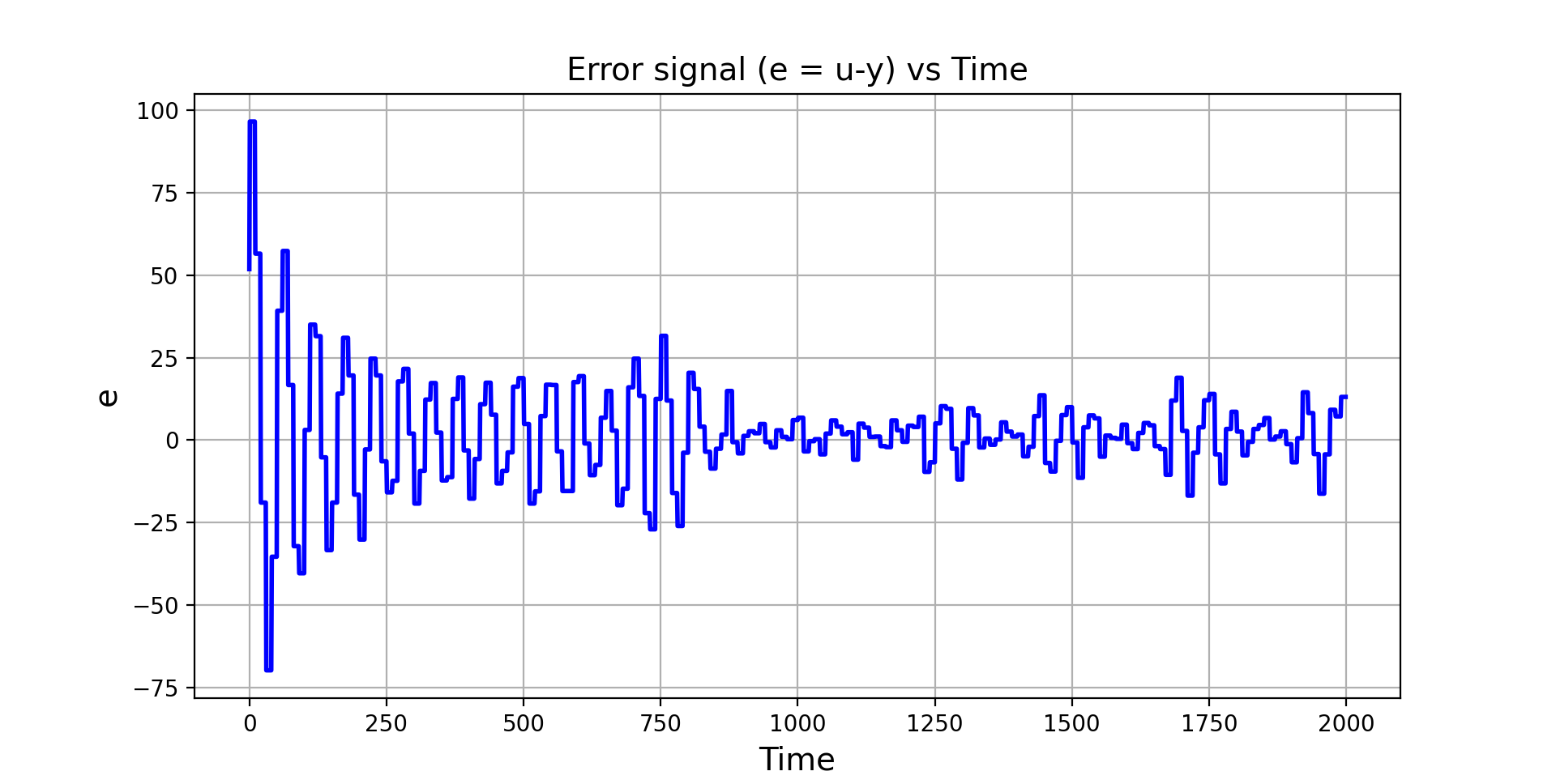}
\caption{The evolution of the error of the pricing control.}
\label{error}
\end{figure}

 The results of the controller test show that the model was appropriately chosen, as the evolution of the error signal tended towards zero when the {\tt fixedDemand} was set to 180. This meant that the target number of people using the incentivised routes was sufficiently met, and the controller could be implemented successfully in the rest of the algorithm (see Figs. \ref{contribution-streets} to \ref{error}). The parameters of the model were also tested in the edge cases of the system; i.e., choosing both high and low values of {\tt fixedDemand}, relative to the population size.

\begin{table*}[!b]
\caption{Simulation evaluation metrics. \label{tab:sim-metrics}}
\centering
\begin{tabular}{|p{15 em}|p{20 em}|p{25 em}|}
\hline
\textbf{Simulation Metric} & \textbf{Question Answered} & \textbf{Benchmark Value} \\ [0.5ex]
\hline
A1: Time taken to reach incentivised route & How effective is the simulation at navigating agents onto an incentivised route once the agent has decided 'yes' to using an incentivised route? & All data points within 2 standard deviations of the mean time.\\
\hline
A2: Simulation density (i.e., number of agents / number of pedestrian allowed lanes) & What is the critical mass of agents needed for the system to be able to function sensibly? & Simulation density $>$ 0.001

(5 users per street; 4616 pedestrian allowed lanes in the map of South Kensington. 5 users were selected as a benchmark after discussions with stakeholders.)
\\
\hline
A3: Simulation density on incentivised routes & Is the system effective at solving the problem of female safety? & Correlation between simulation density on incentivised routes and simulation density $>$ 0.2
\\
\hline
A4: Ratio of agents on incentivised route to agents in simulation & Is the system effective at solving the problem of female safety? & Ratio of pedestrians on incentivised route $>$ 10\%
\\
\hline
\end{tabular}
\end{table*}

\subsection{SUMO Simulation}
\begin{figure}[!t]
\centering\includegraphics[width=0.9\columnwidth]{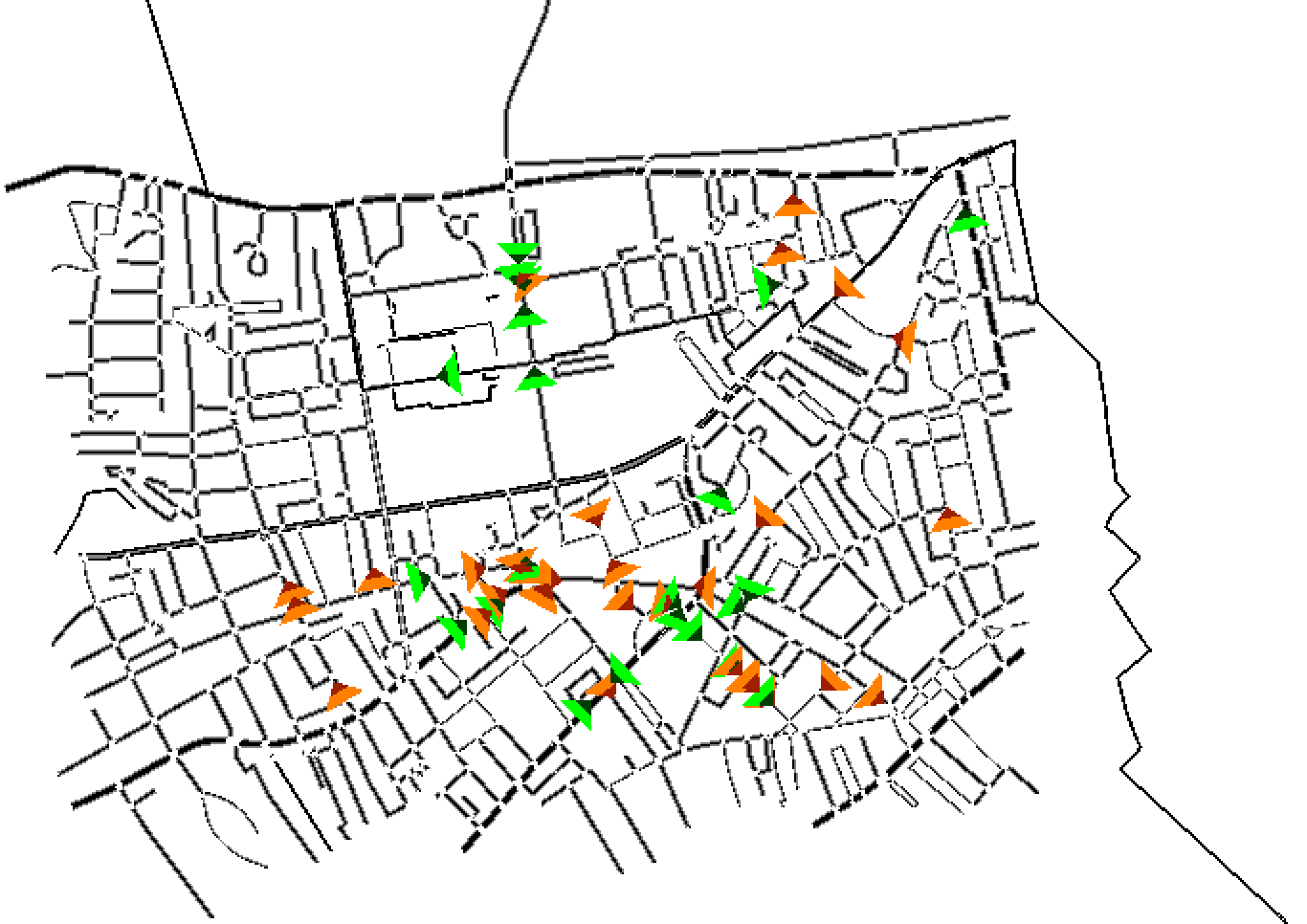}
\caption{If the agents turned green, they had moved onto an incentivised route.}
\label{sim-evaluation}
\end{figure}

Taking into account the three key criteria for success (security, effectiveness and inclusivity), detailed metrics were distilled to measure the validity of the simulation (Table \ref{tab:sim-metrics}).
The test involved running the simulation through 125 cycles, and collecting data that would be analysed to demonstrate whether the test metrics had been met. As the agents navigated onto the incentivised route, their SUMO-GUI visualisation turned green (Fig. \ref{sim-evaluation}), providing visual confirmation of the simulation working as intended. The simulation was tested with several constraints and assumptions as stated below:

\begin{itemize}
  \item number of users would vary between 30 and 50;
  \item each simulation would run for 1000 steps;
  \item the values for the pricing controller used were $\alpha_{c1}$ = -4.01, $\beta_{c1}$ = 0.99 and; $\kappa_{c1}$ = 0.1 \cite{controller-default-values};
  \item and the time interval regarding when a new control signal was sent was set to 10 steps.
\end{itemize}

\begin{figure}[!t]
\centering\includegraphics[width=0.9\columnwidth]{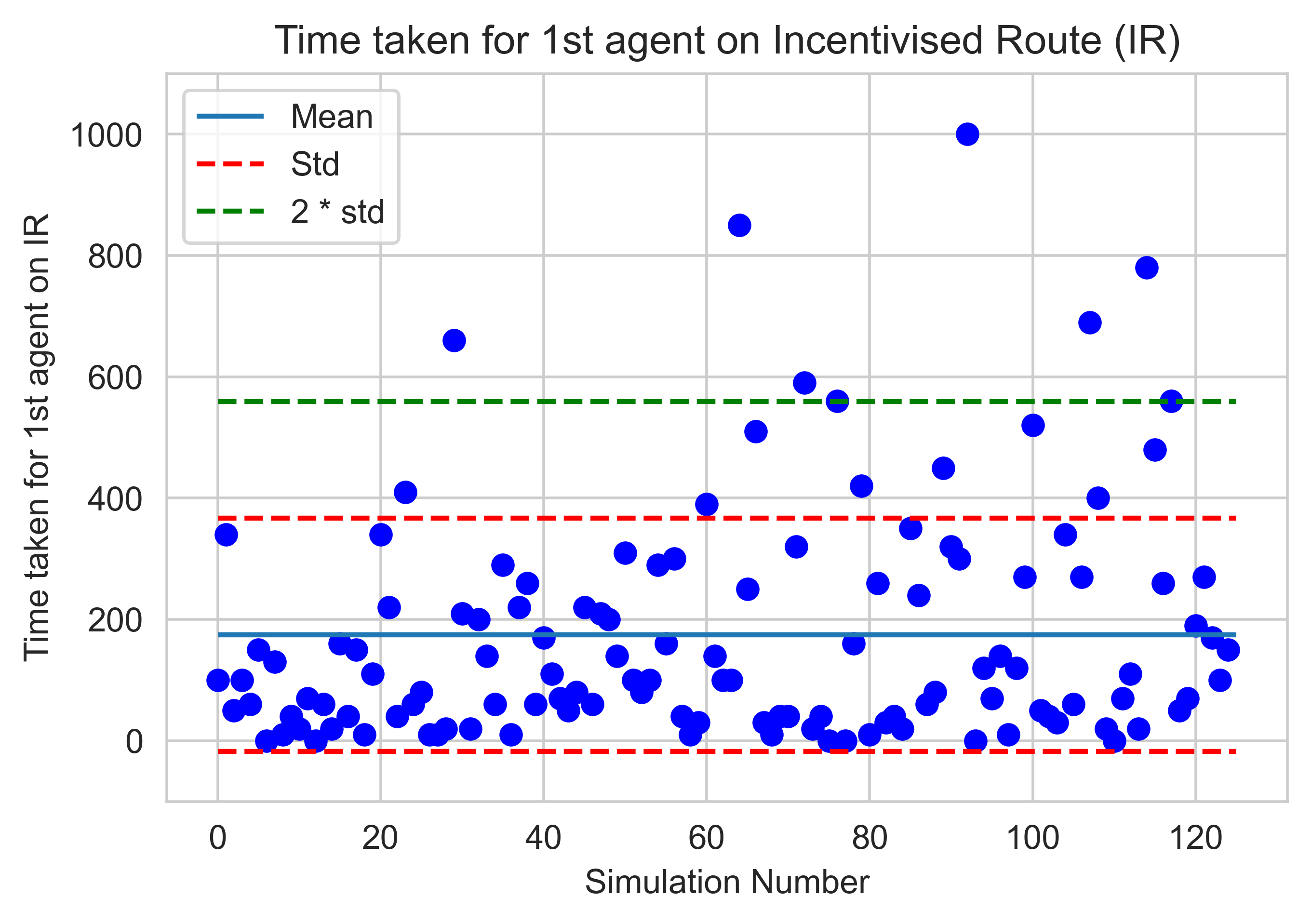}
\caption{Time taken for the first agent to navigate onto an incentivised route (A1, Table \ref{tab:sim-metrics}).}
\label{time-taken-agent-on-IR}
\end{figure}

\begin{figure}[!h]
\centering\includegraphics[width=0.9\columnwidth]{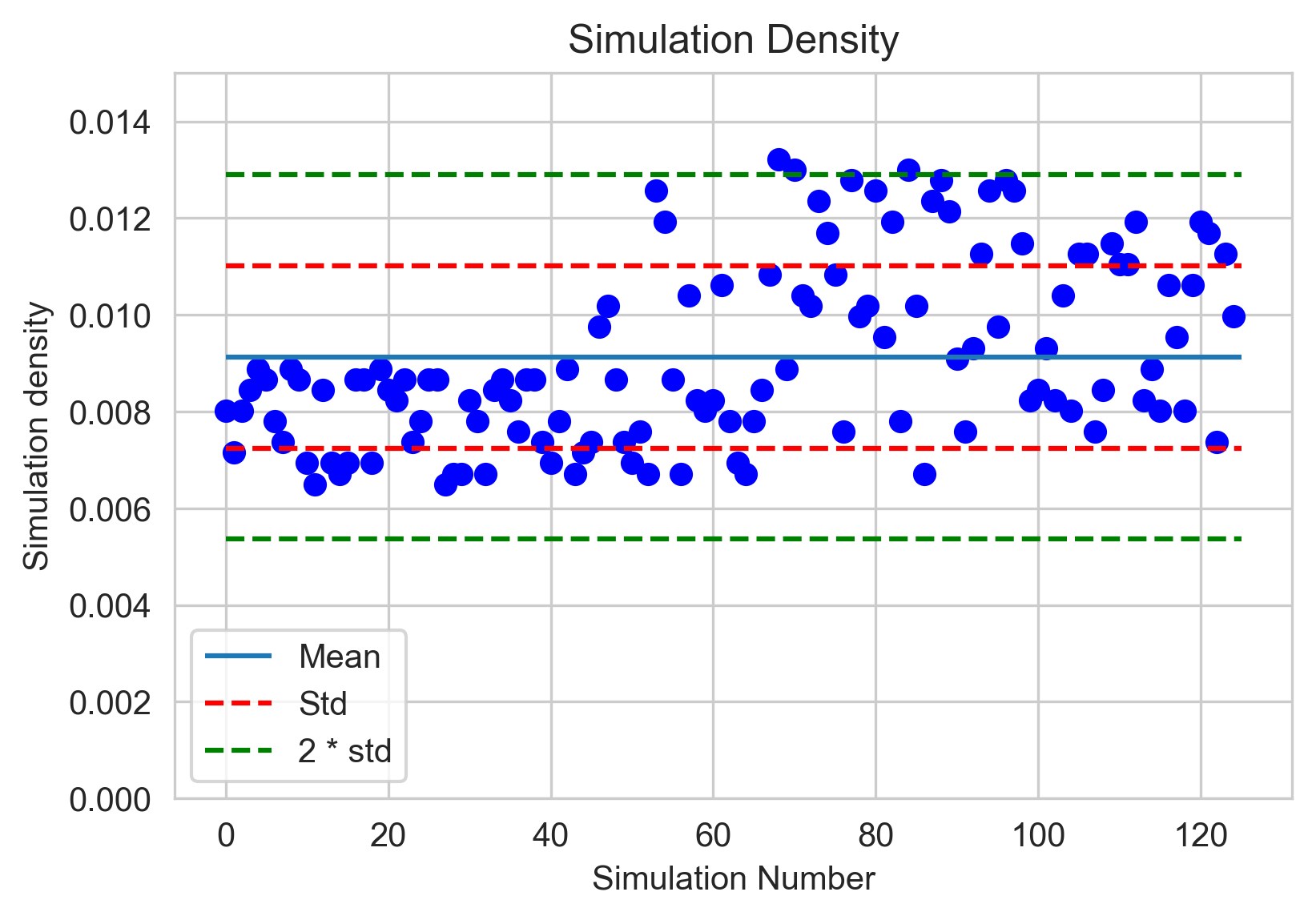}
\caption{Simulation density (A2, Table \ref{tab:sim-metrics}).}
\label{sim-dens}
\end{figure}

\begin{figure}[!h]
\centering\includegraphics[width=0.9
\columnwidth]{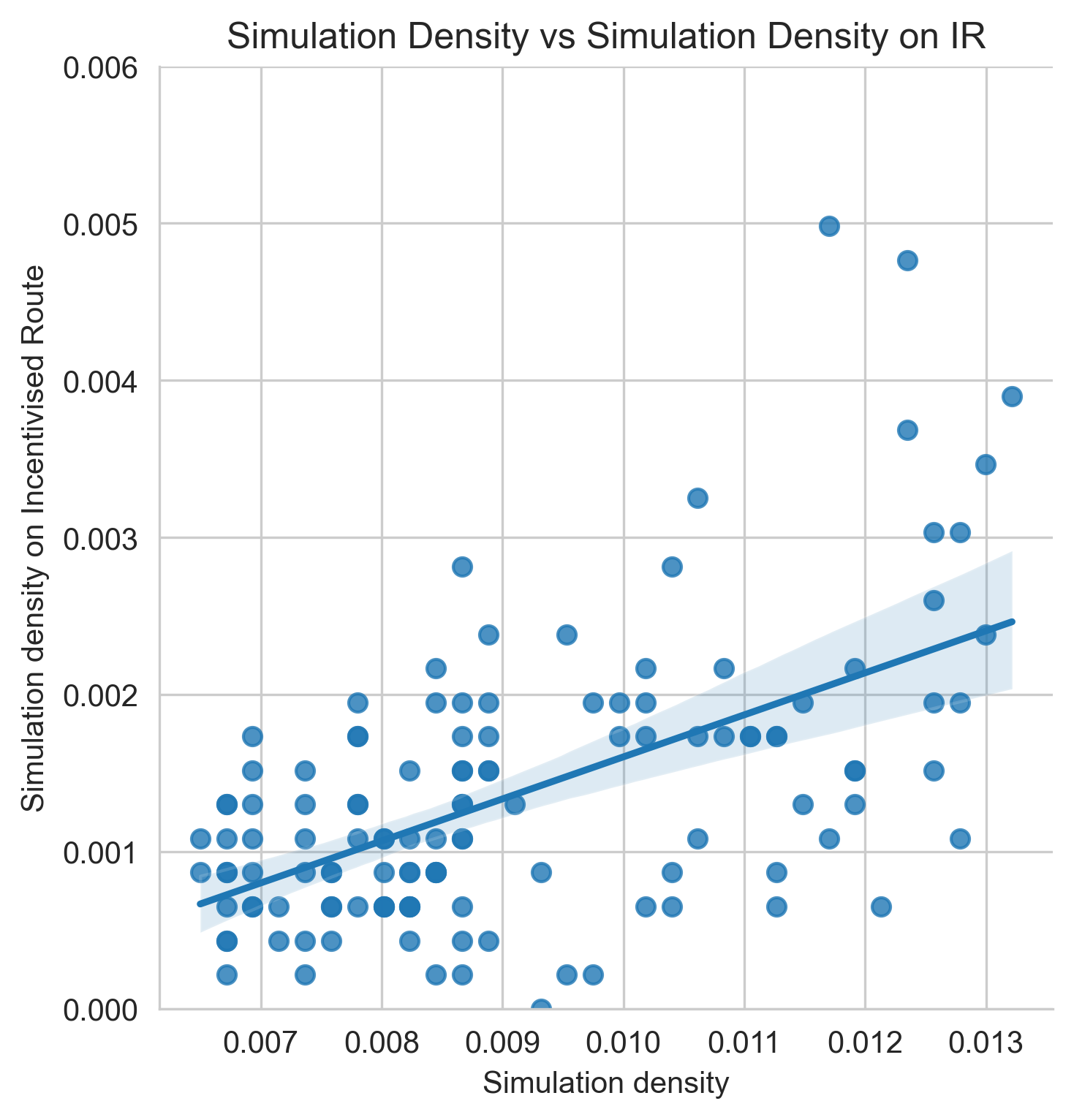}
\caption{Correlation between simulation density and simulation density on incentivised routes (A3, Table \ref{tab:sim-metrics}).}
\label{sim-dens-correlation}
\end{figure}

\begin{figure}[!h]
\centering\includegraphics[width=0.9\columnwidth]{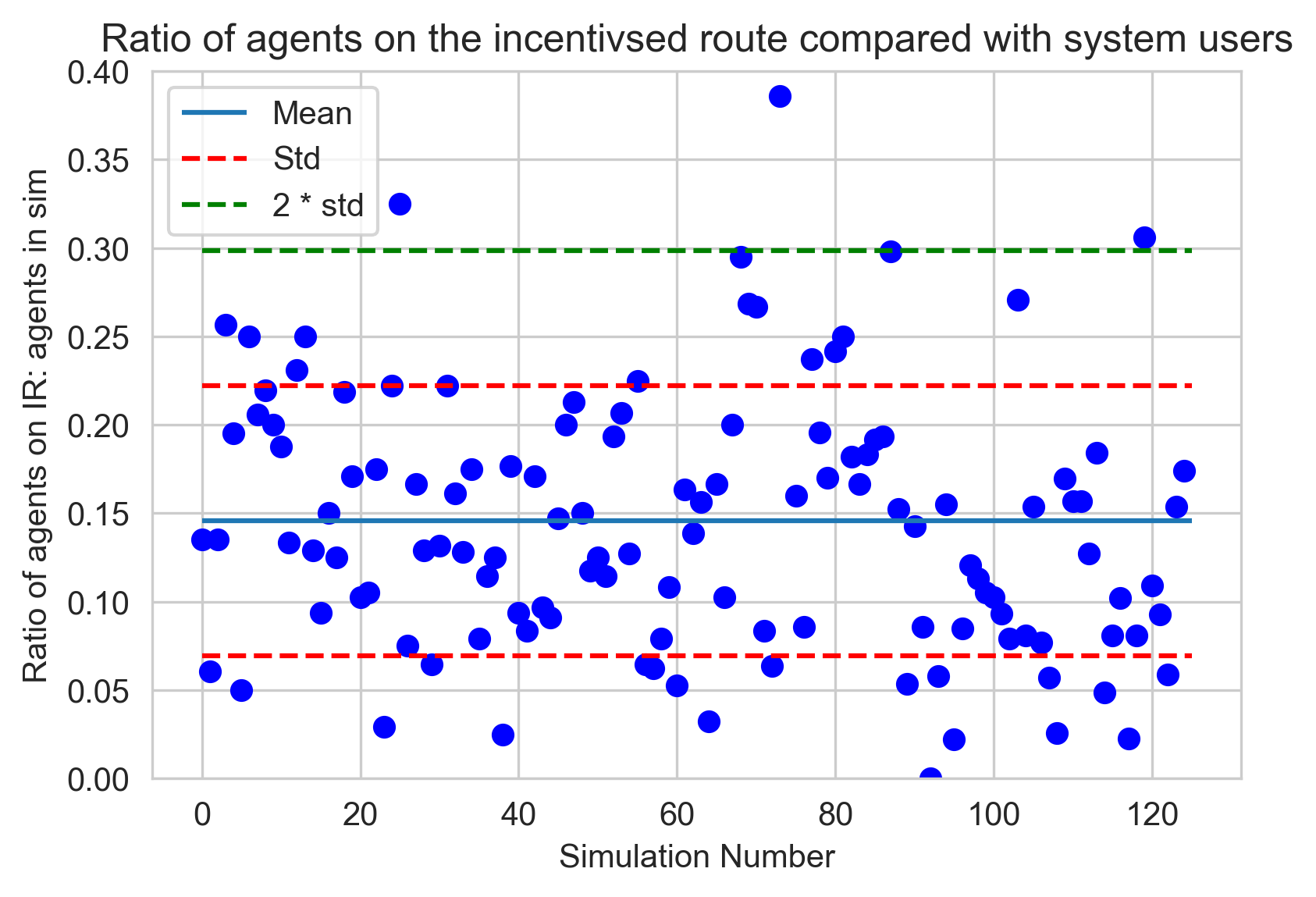}
\caption{Ratio of agents using incentivised routes to all agents (A4, Table \ref{tab:sim-metrics}).}
\label{ratio-of-agents}
\end{figure}

The results of the tests were as follows: (a) Average Time Taken\footnote{That is, average number of simulation steps taken, where each simulation step represents 1s in real time.} for the first agent to navigate onto an incentivised route = 174.64 steps, Standard Deviation = 192.15 steps (Fig. \ref{time-taken-agent-on-IR}); (b) Average Simulation Density = 0.0091, Standard Deviation = 0.0019 (Fig. \ref{sim-dens}); (c) Correlation Coefficient between Simulation Density on Incentivised Routes and Simulation Density = 0.57 (Fig. \ref{sim-dens-correlation}); (d) Average Ratio of Pedestrians Using Incentivised Routes to All Pedestrians = 14.57\%, Standard Deviation = 7.63\% (Fig. \ref{ratio-of-agents}). Fig's. \ref{time-taken-agent-on-IR} to \ref{ratio-of-agents} represent the data graphically, and were used to discern whether the test criteria were met.

Criteria A2, A3 and A4 were achieved during the test. A1 was not met; however, the time taken to reach an incentivised route exceeded the criteria in only 2/125 simulations. There were no simulations, where modelled agents failed to accept a token price; and in only 1 out of 125 simulations (0.8\%) did the simulation fail to navigate at least one agent who wanted to navigate to an incentivised route, to one.

\subsection{Hardware-in-the-loop}
\begin{figure}[!t]
\centering\includegraphics[width=0.9\columnwidth]{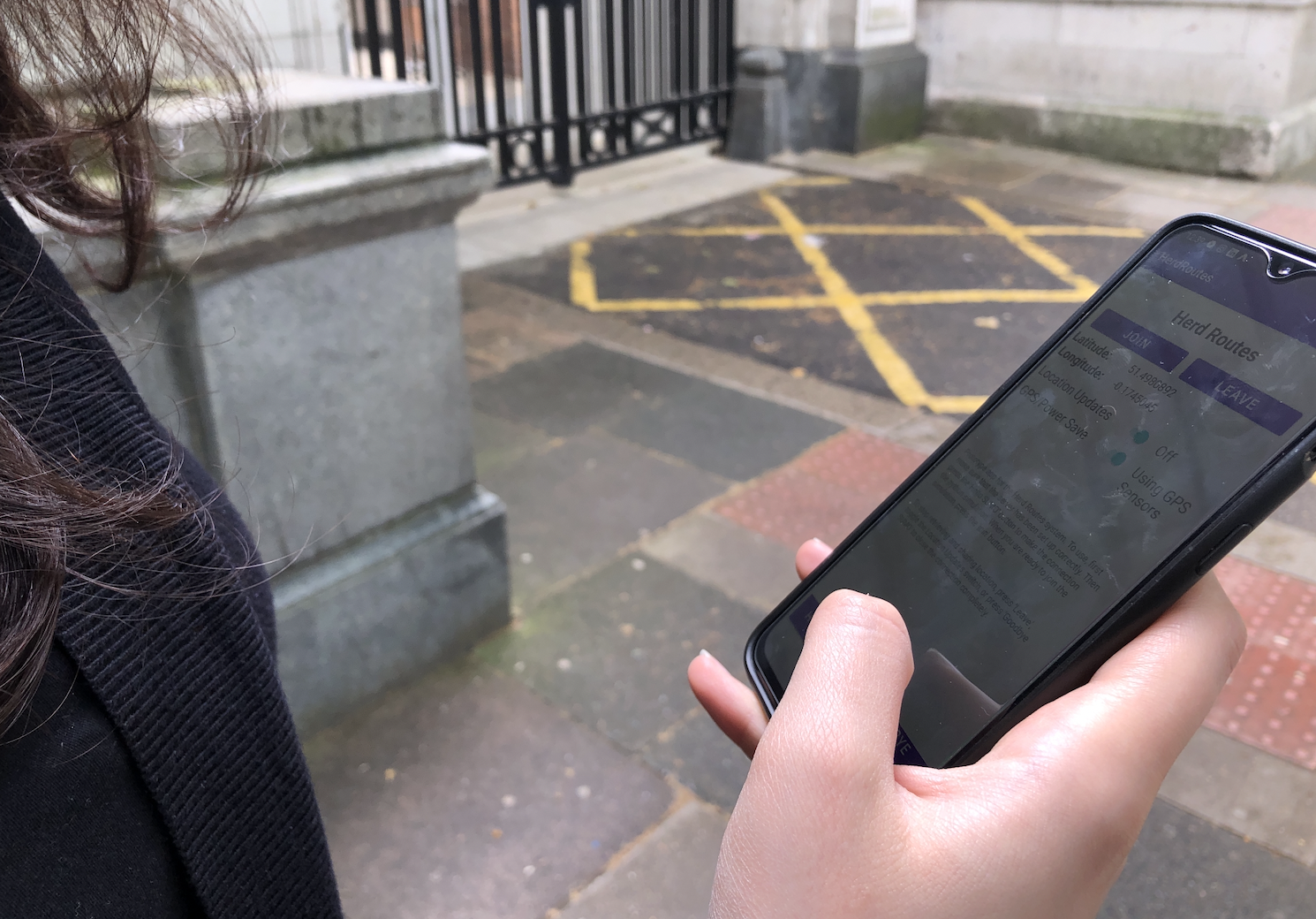}
\caption{Participant using prototype smartphone application to participate in hardware-in-the-loop testing.}
\label{hitl-walk-participant}
\end{figure}

The prototype user interface was loaded correctly and the simulation was successfully connected to and joined in all real human participant cases (Fig. \ref{hitl-walk-participant}). Having completed the test, on average, participants were more likely to be aware of other pedestrians while walking along a route  (as rated on a scale of 1-10). When asked about participating in a large-scale system, feedback included ``it feels safe and makes you feel part of a community''; ``it's a non-invasive way of showing areas which are busier''; ``made me want to use it as I felt I was helping others''; and ``I felt nervous knowing that other people knew where I was''. On the subject of data privacy, participants were on average less happy knowing their location was being used; however, there was a large disparity in opinion. Despite this, 63\% of the participants said that the use of location data was justified within the context of a female safety application.

When discussing the system more generally, the consensus was that a street with 5-10 other pedestrians on it would feel busy and therefore would have a greater perceived safety. The opinion was unanimous that it was acceptable that other system users would benefit from the safety application in the form of tokens. City-specific rewards and social utility tokens were voted as the most desirable option for token forms.

\section{System Considerations}\label{s5}

\subsection{Business Case}
Incentivisation of the specified routes underpins the success of the Herd Routes system. Although this system was designed for maximum social impact, and therefore at its core would be operating as a not-for-profit business, someone would need to provide the value of a token. When exploring who would pay for this system, the following questions were considered:

\begin{enumerate}
  \item What form of token would be exchanged?
  \item Who are the critical stakeholders who would need to sign off on this system in order for it to be rolled out?
  \item Who has a vested interest in this system working?
  \item How does the token exchange work initially, and then over a longer period?
\end{enumerate}

By analysing similar female safety applications, it can be seen that various financing strategies have been used. Some generate revenue from advertisements, where others sell on data-generated insights from their application to third parties (typically government institutions, local councils, and urban planning companies). All instances agree on the fact that the user in question should not have to pay. The options for the system financing are discussed in Table \ref{tab:system-financing}; however, from the qualitative feedback received, options with tokens based on social utility were preferred.

\begin{table*}[!t]
\caption{Options for Herd Routes System Financing
\label{tab:system-financing}}
\centering
\begin{tabular}{|p{10 em}|p{17.5 em}|p{12.5 em}|p{17.5 em}|}
\hline
\textbf{Institution / Organisation} & \textbf{Rationale} & \textbf{Advantages} & \textbf{Disadvantages}\\ [0.5ex]
\hline
Local shop fronts
& Herd Routes generate traffic flow along the routes. High-street shops will benefit from this service and so should pay.
& Consistent revenue stream, benefiting the local economy in multiple ways.
& Opens up the possibility of a conflict of interests between the route selectors and the system users. Becomes too predictable, could lead to system misuse. \\
\hline
Cryptocurrencies / Ledgers
& Use the system to bootstrap ledgers such as IOTA, increasing their popularity and getting them closer to the required number of network users; e.g., NFT Shoe, SweatCoin \cite{sweatcoin}.
& Short-term lucrative business model.
& Excludes potential users that would not have use for cryptocurrencies / don't have interest or see value in it.
\\
\hline
Local Councils / London Boroughs
& Local government councils have invested interest in creating as safe an environment as possible; e.g., Path Community \cite{path-community}.
& Lends system credibility, and speeds up implementation times.
& Varies borough to borough. Would be difficult to coordinate across multiple local councils initially
\\
\hline
Big corporations / institutions on Women's Safety Charter London \cite{womens-safety-charter}
& Use CSR policies to fund a social impact project.
Demand from employees and customers.
& Access to a large source of financing.
& Aligns the system with goals that may deviate from the original mission.
\\
\hline
Mayor of London
& Local government councils have an interest to create as safe an environment as possible. Tokens could be in the form of city-wide valued commodities such as free TFL journeys.

& Lends system credibility, and speeds up implementation times. Opens up potential for highly valued societal tokens.

& Potential cap on the value of the tokens.

System may be vulnerable to whether priority or funding changes within the public sector.
\\
\hline
Home Office \& Police
& Shared common objective of reducing crime in public spaces.

& Strong alignment of objectives that will not change in the future.

& Privacy concerns, and vulnerable to changes in public sector funding.
\\
\hline
\end{tabular}
\end{table*}

\subsection{Security}
There is general concern about the movement toward smart cities and the lack of data security. A specific concern is the ``lack of opportunity for giving meaningful consent to processing of personal data'', as outlined in \cite{privacy-smartcities}. This system is voluntary, and upon system onboarding, the user grants permission for their location and system identification number to be sent in an encrypted format to the Tangle. By the very nature of the ledger, all transactions from and to it are traceable. This means that any data leak or hacking of the encrypted identifiable information generated by the Herd Routes system would allow the ledger to be traced for all past and future messages or transactions linked to the entity tied to that transaction. This would quickly identify the entity that had misused the system, and would be reported and dealt with accordingly.

Throughout the wider IoT industry and research community, there is a debate on the value of data privacy versus the potential for social utility. During the Consequence Scanning workshop that was conducted, it was concluded that all four interviewees already used location-based tracking services, such as Find my Friends, or Snap Maps. From user interview responses, it was clear that there was an acknowledged trade-off between maintaining privacy, especially with other indirect stakeholders such as parents, friends, or the authorities, and making sure that the primary functionality of the safety system worked as well as it could, maximising protection of their welfare.

\textbf{Preventing System Misuse:} The Herd Routes system is designed so that all citizens can use it, but vulnerable pedestrians benefit the most from the creation of safer public spaces. As mentioned previously, this is reliant on location sharing and storing of all system users in an encrypted format. However, to a malevolent system user, Herd Routes could be seen as a search for female pedestrians' locations. Design features could be integrated to encourage proper use, such as the sharing of locations of all users, including any malevolent system user and employing a `referral system' for new users joining the system. When new users to the system are referred, a chain of user referrals is created. If one user within the chain misused the system, then all of the users linked to that person would also be removed from the application. Adopting a feature like this would mean that the risk of system misuse would decrease, as it is not in the interest of benevolent system users to refer new people who may misuse the system.

Ride-hailing apps such as Uber\footnote{\url{https://www.uber.com/gb/en/}} all face similar issues regarding system misuse and threat to user safety. In these cases, the success of the application relies on the ``assurance to deliver what is expected and to maintain the trust promised in the mission of the organisation'', as stated in \cite{uber-passenger-safety}. By adding safety features to the application, such as sharing a trip with trusted contacts, real time ID check for drivers, and verifying a trip by using a unique PIN, the risk of system misuse is mitigated.

\section{Conclusion \& Future Work}\label{s6}

This paper proposed a novel solution to the pressing issue of improving the safety of female pedestrians in public spaces. By developing an incentivisation algorithm, which utilised a dynamic pricing controller and integrated with the IOTA Tangle, a proof-of-concept was tested. Hardware-in-the-loop testing allowed real human behaviours to be incorporated into the simulation, as well as gaining user validation for the concept. The results indicate that this system has potential to have real impact not only by improving female pedestrian safety in the short term by generating busier streets to walk down, but also having longer-term impact on changing the way that all citizens regard the societal behavioural change required to minimise gender-based violence in public spaces. Directions for future work include: (i) Incorporating a more intelligent method of selecting which routes are incentivised. A smarter algorithm could be developed by using environmental APIs, the I3 database or using proximity of current system users as an input to the incentivised route generation; (ii) Prototyping the token exchange and developing the stakeholder network to validate financing of the system. A `token calculation' algorithm would need to be designed. This would calculate how many tokens priced at that value should be transferred to each user based on time spent on the route; and finally, (iii) Developing the prototype smartphone application so it has map navigation functionality, and visibility over other system users.

\end{document}